# Magnetic-free terahertz nonreciprocity via temporal dissipative barriers

Mingyu Tong[1,3,4,5,†], Yuze Hu[1,2,†], Siyang Hu[6,†], Hongsheng Chen[1,3,4,5,*],
Tian Jiang[2,*], Yihao Yang[1,3,4,*]

[1] Interdisciplinary Center for Quantum Information, State Key Laboratory of Extreme Photonics and Instrumentation, ZJU-Hangzhou Global Scientific and Technological Innovation Center, Zhejiang University, Hangzhou 310027, China.
[2] Institute for Quantum Science and Technology, College of Science, National University of Defense Technology, Changsha, 410073, China
[3] International Joint Innovation Center, The Electromagnetics Academy at Zhejiang University, Zhejiang University, Haining 314400, China.
[4] Key Lab. of Advanced Micro/Nano Electronic Devices & Smart Systems of Zhejiang, Jinhua Institute of Zhejiang University, Zhejiang University, Jinhua 321099, China.
[5] Shaoxing Institute of Zhejiang University, Zhejiang University, Shaoxing 312000, China.
[6] College of Advanced Interdisciplinary Studies, National University of Defense Technology, Changsha, 410073, China.

[†]These authors contributed equally to this work.
[*]Corresponding author. Email: yangyihao@zju.edu.cn (Y.Y.); tjiang@nudt.edu.cn (T.J.); hansomchen@zju.edu.cn (H.C.)

## Abstract

Terahertz (THz) nonreciprocal devices are essential for advancing future fundamental science, wireless communications, imaging, and sensing. Current THz nonreciprocal devices mostly rely on magnetic materials, which, however, suffer from large volume, operation under an external magnetic field, and low-temperature environment, rendering them poorly compatible with miniaturized developments. Here, we propose an unconventional method for achieving THz nonreciprocity free from magnetic materials. The scheme relies on a temporal dissipative barrier, a transient loss variation generated by photoexcited carriers, and the nonreciprocity arises from the distinct coupling behavior for different polarizations with the barrier. The isolation efficiency correlates with the temporal barrier width, resonant mode detuning, and the working frequency, and has been significantly enhanced by introducing a dark mode. We experimentally confirm our method in a THz optically active metasurface with wave-flow isolation exceeding 20 dB across a bandwidth greater than 0.4 THz. Theoretical predictions indicate peak isolation surpassing 60 dB, with experimental results achieving over 30 dB at 0.7 THz. Our approach unlocks the potential of miniaturized, integrated, magnetic-free THz nonreciprocal devices for various applications.

Terahertz (THz) wave holds promise for unlocking terabit-per-second wireless connections, which are pivotal for future 6G communications[1, 2, 3, 4], and it has impactful applications in imaging[5, 6, 7, 8, 9], sensing[10, 11, 12, 13, 14], and beyond[15, 16]. Nonreciprocal THz devices, such as isolators and circulators, can protect THz sources, mitigate multipath interference, and enable stable communication or operation[17, 18, 19, 20]. Conventional THz nonreciprocal devices are realized through the Faraday effect and the magneto-optical Kerr effect[20, 21, 22], or by harnessing the inherent nonreciprocal directional dichroism in magnetic materials[23, 24, 25]. However, the weak magneto-optical effects are less conducive to efficient operation at terahertz frequencies. Moreover, these devices rely on external magnetic fields and require long interaction lengths, which are too bulky to integrate into modern miniaturized photonic systems[26].

Temporal modulation can break Lorentz reciprocity, fulfilling the criteria for magnetic-free nonreciprocity and isolation[19, 27]. Nonreciprocal devices based on temporal modulation have been demonstrated in many systems, such as acoustics[28], microwaves[29], and optics[30]. As an elementary operation in temporal modulation, temporal boundaries have garnered significant attention owing to their critical roles in many intriguing phenomena, including time reflection and refraction[31, 32, 33], coherent wave control[34], frequency conversion[35, 36, 37], photon acceleration[38], non-Abelian Aharonov-Bohm interference[39], and photonic cavity engineering[40]. The boundaries act as a time-domain equivalent to spatial boundaries between different media; they occur when a system experiences an abrupt change in material properties over time while remaining uniform in space, with a timescale comparable to the oscillation period of the electromagnetic wave. Although nonreciprocity based on a single time boundary has been theoretically proposed, its realization still faces challenges[41]. Active metamaterials, which control the morphological transformation of conductive resonant structures and their mode field profiles using electrically or optically tunable natural materials, are ideal for realizing temporal boundaries[42]. Particularly, the ultrafast manipulation of optical resonances in a time-variant medium can obtain switching speeds of the coupling coefficient that matches the oscillation period of THz waves[36, 37, 43, 44]. To date, however, exploiting this approach to achieve THz nonreciprocity and isolation has never been explored.

Here we propose and experimentally demonstrate the magnetic-free THz nonreciprocity in a metasurface with a single temporal dissipative barrier. Rather than relying on spatiotemporal modulation, our method suddenly breaks the coherent cancellation between modes in a polarization-dependent photonic system by abruptly changing its electromagnetic properties uniformly in space, realizing a temporal barrier (TB) capable of inducing extremely fast phase shifts between two resonant modes. The transient phase shift disrupts the coherent cancellation between resonant modes, thereby enabling polarization conversion. In contrast, the coupling effect in the reverse direction is counterbalanced by the external field, even when the TB is in effect. The measured isolation ratio reaches up to 30 dB at 0.7 THz and can be conveniently controlled by adjusting the pump fluence. Our results introduce the concept of TB to nonreciprocal THz devices, transcending the limitations of conventional magnetic materials and offering opportunities for constructing miniaturized, integrated, magnetic-free nonreciprocal devices.

**Results**

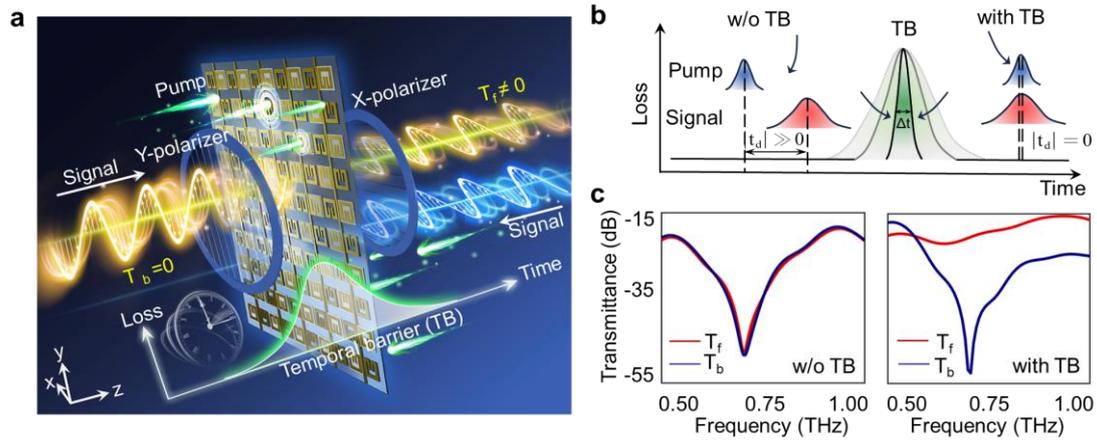

**Fig. 1 | Nonreciprocal THz transmission in an optically tunable metasurface with a TB. a**, Schematic illustration of THz nonreciprocal transmission, where the forward incident y-polarized signal is converted into an x-polarized wave (yellow line), while the inverse process is prohibited (blue line). **b**, Illustration of the TB. The TB can turn on or off, depending on the time delay ($t_d$) between the arrival times of the pump and signal pulse peaks. The left 'without (w/o) TB' represents the regime where the pump wave is far before or after the input signal; the right 'with TB' is where the TB disturbs the original signals driven by the pump pulse. The middle Gaussian pulses with varied full width at half maxima (FWHM) represent the TB in different widths. **c**, Experimentally measured transmittance for forward ($T_f$) and backward ($T_b$) incidence through the metasurfaces without (left) and with (right) a TB, respectively.

The nonreciprocal THz device is achieved with a TB-based metasurface, schematically displayed in Fig. 1a. The metasurface consists of an array of split-ring resonators (SRRs) exhibiting two classical inductance-capacitance (LC) resonances, a plasmonic mode supported by a continuous long stick, and a dark mode that does not couple to the radiation. To break the static state and establish a temporal boundary, a femtosecond laser excitation is used to induce inter-band transitions and a surging free carrier density in germanium (Ge). When the conductivity of Ge film in the gap of SRRs suddenly increases, it causes an ultrafast rise in Ohmic losses that suppresses one of the resonance modes. The ultrafast rise and fall of the nonradiative losses are defined here as the TB.

The time delay ($t_d$) signifies the interval between the arrival of the input THz signals and the pump pulses at the metasurface. The response of the TB metasurface can be categorized into two operational regimes depending on the time delay, as shown in Fig. 1b. Here, 'w/o TB' indicates that the input pulse reaches the metasurface far behind or ahead of the pump pulse, meaning that the modes driven by this input pulse are undisturbed; thus, the responses reflect a stationary state. However, when the pump pulse synchronizes with the input pulse upon arrival at the metasurface, more intriguing temporal dynamics emerge, as the operation of the metasurface transcends simple steady-state responses. Intuitively, nonreciprocal polarization conversion arises from this dynamic mode variation process at the TB. In our work, the metasurface in its equilibrium state displays identical transmittance for both forward and backward incidence due to time-reversal symmetry, as shown in Fig. 1c. However, with the influence of the TB, isolation exceeding 20 dB is achieved across 0.4 THz, with a maximum isolation of 30 dB at 0.7 THz.

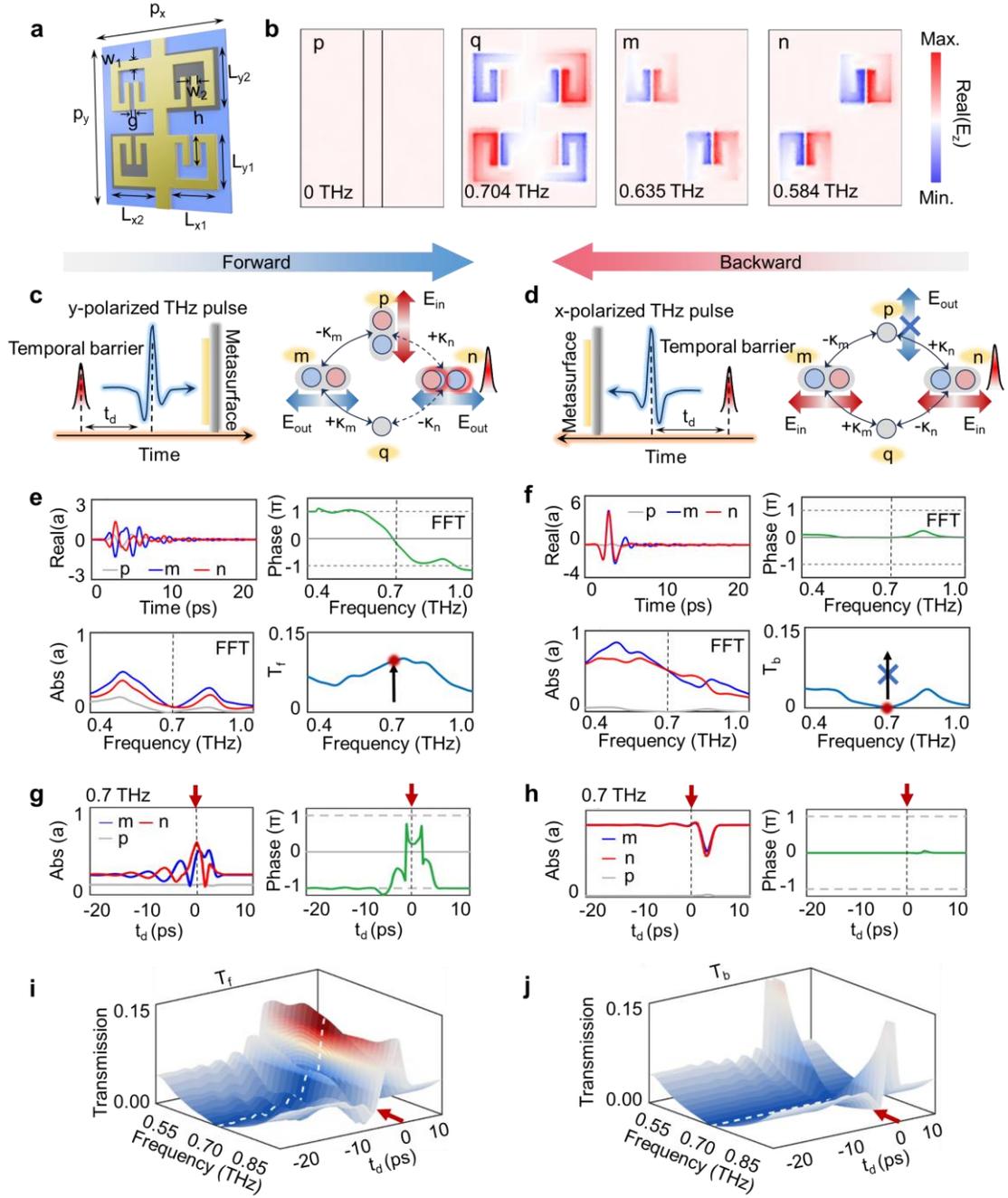

**Fig. 2 | Metasurface design and theory of the nonreciprocal polarization conversion based on a TB. a**, Schemic of the metasurface structure. Photocarrier excitation occurs only in the Ge film (gray region). The geometric parameters are $p_x = 100\,\mu m$, $p_y = 140\,\mu m$, $L_{x1} = 37\,\mu m$, $L_{x2} = 41\,\mu m$, $L_{y1} = 37\,\mu m$, $L_{y2} = 45.5\,\mu m$, $w_1 = 6\,\mu m$, $w_2 = 4\,\mu m$, $g = 4\,\mu m$, and $h = 24\,\mu m$, respectively. **b**, Eigenmode distribution of four modes: a y-polarized plasma mode 'p', a dark mode 'q', and two x-polarized modes 'm' and 'n'. **c,d**, Schematic illustration of polarization-dependent THz transmission and simplified model for the forward and backward processes with a TB. **e,f**, Real parts of the bright modes evolving over time (top-left) and their amplitudes distribution in the frequency domain (bottom-left), the phase difference between modes 'm' and 'n' in the frequency domain (top-right), and theoretical results for the transmission (bottom-right) of forward

and backward incident THz signals with a TB ($t_d = 0$ ). **g,h,** Amplitudes of three bright modes and the phase difference between modes 'm' and 'n' at 0.7 THz evolve with different $t_d$ for forward and backward incident THz signals, as indicated by the white dashed lines in **i** and **j**. **i,j,** Theoretical results for transmissions $T_f$ and $T_b$ as functions of time delay and frequency. The red arrows indicate the moments of maximal nonreciprocal isolation ratio.

We designed a semiconductor-metal hybrid metasurface, as illustrated in Fig. 2a. The proposed unit cell of the metasurface consists of two pairs of SRRs oscillating in the x-direction, with their resonating frequencies splitting due to the introduction of the semiconductor film, marked as 'm' and 'n' modes in Fig. 2b, respectively. One continuous metal stick oscillating along the y-direction (working at 0 THz) couples these two sets of SRRs, denoted as 'p'. A dark mode, labeled as 'q', is supported by the interaction of four SRR scatterers and plays a significant role in the isolation ratio (see details in Supplementary Note).

The dynamics of the mode evolution during the polarization conversion process are comprehensively understood with the assistance of temporal coupled mode theory (TCMT). We model the resonators as a four-mode two-port system under external incidence (see details in Supplementary Note). Here, three bright modes (with resonant frequencies $f_m$, $f_n$, $f_p$ and radiation/absorption decay rates $\gamma_m$, $\gamma_n$, $\gamma_p$, $\gamma'_m$, $\gamma'_n$, $\gamma'_p$) interact with a dark mode (with resonant frequency $f_q$ and radiation/absorption decay rate $\gamma_q$, $\gamma'_q$). External stimuli such as visible light pulses can induce transitions from insulating to conducting states in Ge films, increasing the nonradiative loss of the Ge-incorporated 'n' mode and suppressing its resonance, leading to an abrupt decoupling from other modes. We model this effect using a time-varying loss function $\gamma'_n(t)$, which changes within a few femtoseconds before returning to its initial state, allowing the temporal variation to be approximated by a Gaussian pulse. The couplings between the modes are described by $\kappa_{qm} = -\kappa_{pm} = \kappa_m$ and $\kappa_{pn} = -\kappa_{qn} = \kappa_n$, respectively. Defining $|s^i\rangle = [s^i_{1x}, s^i_{1y}, s^i_{2x}, s^i_{2y}]^T$ ($i = +, -$) as the complex amplitudes of the incoming and outgoing waves at the first and second ports, we can describe the scattering process for the forward and backward transmittance by:

$$T = \begin{pmatrix} T_f \\ T_b \end{pmatrix} = \begin{vmatrix} s^-_{2x}/s^+_{1y} \\ s^-_{1y}/s^+_{2x} \end{vmatrix}.$$

Based on the constructed TCMT, we investigate the essential ingredients for nonreciprocity.

Figure 2c-j illustrate the possible paths for nonreciprocal polarization conversion. The distribution of three bright modes ('p', 'm', 'n') is compared when driven by a y-polarized (for forward propagation) and x-polarized (for backward propagation) THz pulse, respectively. For forward propagation, depicted in Fig. 2c,e, a y-polarized

incident wave stimulates the 'p' mode, concurrently exciting the x-polarized modes 'm' and 'n' out of phase due to opposing coupling constants. Introducing a TB abruptly elevates the nonradiative loss of mode 'n', leading to associated variations among resonators in the system. Figure 2e is a typical case with $t_d = 0$, where the energy within resonator 'n' is diminished both in the time domain (top-left) and the frequency domain (bottom-left). The phase difference between 'm' and 'n' shifts from $\pi$ (top-right), finally leading to the emission of an x-polarized electric field (bottom right) in Fig. 2e. However, under the driven of an x-polarized external field for backward propagation, the 'm' and 'n' modes oscillate in phase with similar amplitudes, as shown in Fig. 2d,f. The opposite coupling coefficients, $\kappa_{pm}$ and $\kappa_{pn}$, counterbalance the coupling effect on the 'p' mode, thereby preventing the conversion of the y-polarized wave. Taking the case $t_d = 0$ in Fig. 2f as an example, although the introduction of a TB diminishes the amplitude of the 'n' mode (top-left) and converts part of its energy to the 'p' mode (bottom-left), the phase difference between modes 'm' and 'n' is preserved due to the driven of external x-polarized field (top-right). The non-radiative state of the 'p' mode is sustained around frequency $f_q$ (0.7 THz) and results in a forbidden polarization conversion (bottom-right). Thus, the TB contributes solely to the polarization conversion from y-polarized to x-polarized, enabling unidirectional photonic polarization conversion.

A comprehensive exploration of THz transmission as a function of time delay in the frequency domain is presented in Fig. 2i,j for both forward and backward propagation. Away from the TB (indicated by red arrows), reciprocity is maintained, while a significant nonreciprocal transmission behavior can be observed at the TB. The amplitudes and phase evolution of resonators at 0.7 THz, where backward transmission vanishes while forward transmission remains non-zero, are illustrated in Fig. 2g,h, as a typical case for nonreciprocal transmission. For forward propagation, when $|t_d| \gg 0$, no x-polarized radiation occurs because the 'm' and 'n' modes oscillate in opposite phases with approximately equal amplitudes. However, as $|t_d|$ decreasing, the TB starts to interact with the tail end of the THz pulses, causing a small perturbation in the amplitudes of modes 'm' and 'n'. As $|t_d|$ approaches zero, the oscillation strengthens due to better overlap between the THz pulse and the TB, shifting the phase difference between modes 'm' and 'n' from $\pi$ to $0.2\pi$ at $t_d = 0$, thereby resulting in the maximum radiation of the y-polarized wave. For backward propagation, no y-polarized radiation occurs due to the in-phase oscillation of modes 'm' and 'n' with opposite coupling constants. The phase difference is maintained at zero by the external x-polarized field, and the coupling from modes 'm' and 'n' to the 'p' mode is counteracted all the time. Consequently, we can achieve a polarization-dependent THz nonreciprocal transmission through a TB.

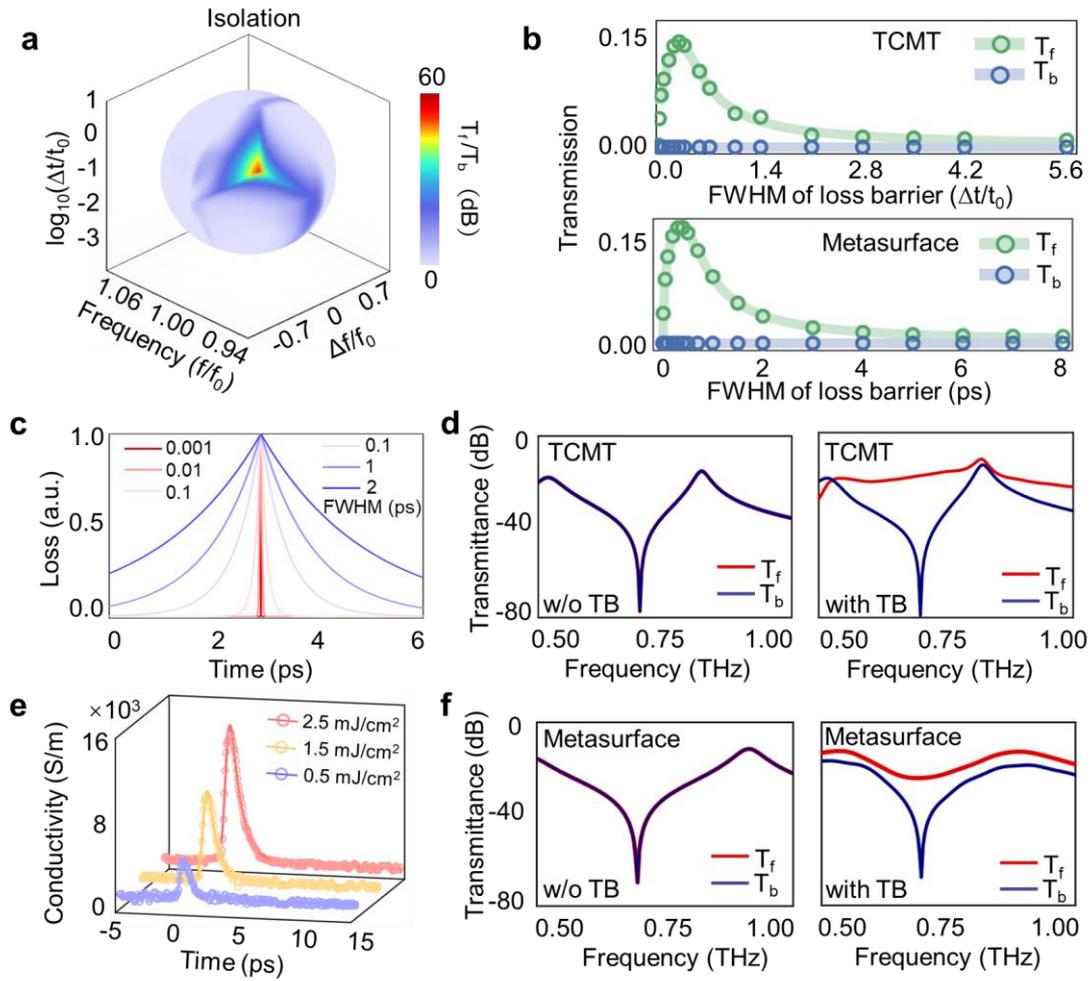

**Fig. 3 | Optimal nonreciprocity of the metasurface induced by the temporal barrier. a**, Comprehensive diagram based on the TCMT model, showcasing the isolation ratio ($T_f/T_b$) as a function of the detuning of two SRRs ($\Delta f = f_m - f_n$), FWHM of the TB ($\Delta t$), and the working frequencies (f). **b**, Gaussian pulses in different FWHM as a TB in the analysis of TCMT and simulations of the metasurfaces. **c**, Extracted transmittance from TCMT (top) and metasurfaces simulation (bottom) at $f_0 = f_q (t_0 = 1/f_0)$. Dots denote the calculated data and curves are fitted for better visualization. **d**, Theoretical results of the forward and backward transmittance without (left) and with (right) TB, with the FWHM of TB setting as 0.28 $t_0$. **e**, Time-resolved photocarrier recombination dynamics within a 200-nm-thick layer of Ge film. This is deduced by the negative differential transmission ($-\Delta E/E_0$) of the THz wave, attributed to the photoinduced conductivity of the semiconductor layer. The data demonstrate a complete recovery time of less than 5 ps. **f**, Simulation results without (left) and with (right) the photoexcitation of Ge as the TB.

Based on the TCMT analysis, we have obtained a diagram of the isolation ratio as a function of the FWHM of the TB, SRR detuning, and operational frequencies, as depicted in Fig. 3a. Obviously, the maximal isolation is at the center of the 3D parameter space, featuring no detuning between the two SRRs, $0.28t_0$ for the FWHM of the TB, and the working frequency aligns with the position of the dark mode. In the metasurface simulation within the THz region, the FWHM of the TB varies from $10^{-3}$ ps to 8 ps (see Fig. 3c). The spectral response is highly sensitive to the width of the TB, as shown in the bottom panel of Fig. 3b. The forward transmittance undergoes a sharp rise, followed by a relatively gradual decline, while the backward transmittance remains insignificant. These observations are consistent with the TCMT results in the upper panel of Fig. 3b. For effective temporal manipulation, an appropriate FWHM is imperative since the flat components in the metasurface can only afford very short interaction lengths and times.

To provide a clear perspective on the spectral response of the designed metasurface, the theoretical results for THz transmittance at the FWHM setting of $0.28\ t_0$ are presented in Fig. 3d without (left) and with (right) TB. With the TB in place, an isolation level approaching 60 dB is attainable, whereas without the TB, the result is completely reciprocal. Subsequently, to explore practically feasible conditions for nonreciprocal conversion within the THz spectrum, we investigate the temporal carrier evolution of Ge under various pump fluences. As depicted in Fig. 3e, the photoexcitation and recombination of Ge exhibit little asymmetric in their distribution. Integrating a realistic TB from Ge into the simulation, as illustrated in Fig. 3f, a pronounced nonreciprocal behavior for THz transmission (right) can be observed, sharply contrasting with the static state (left) and highly consistent with the theoretical results in Fig. 3d.

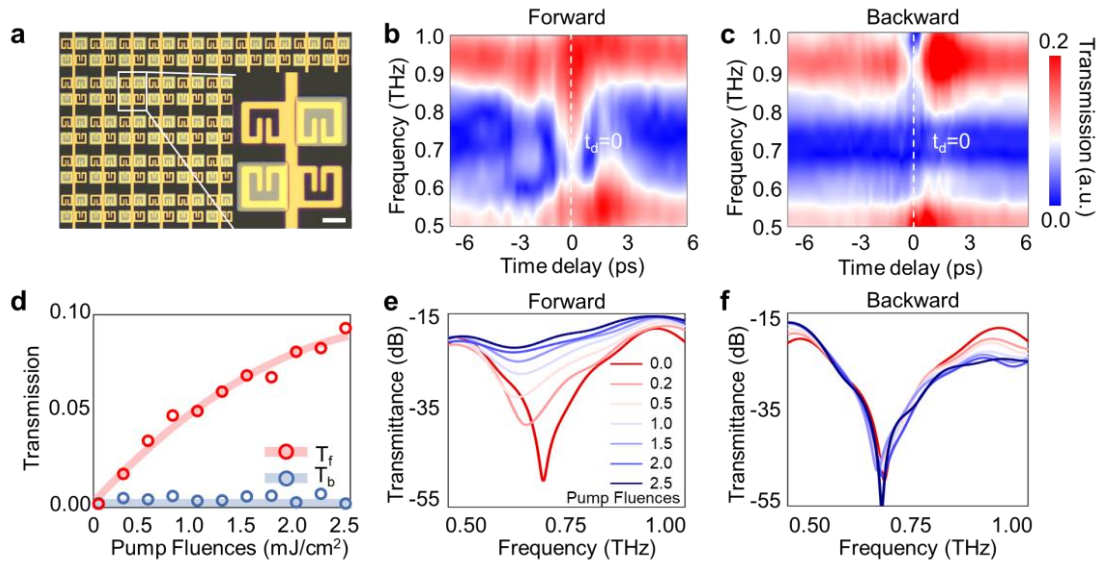

**Fig. 4 | Experimental observation of magnetic-free THz nonreciprocity. a**, Optical microscopy images of the fabricated metasurfaces and single unit-cell. Scale bar, 20 μm. **b,c**, 2D plot showing the measured forward and backward output transmittance spectra as a function of the pump-probe delay $t_d$, at an excitation fluence of 2.5 mJ/cm². The white dashed line indicates the position of the TB. **d**, Relation between the maximum transmission and input power at $t_d = 0$. The measured data are plotted with circles and fitted with a solid line. **e,f**, Experimentally measured forward and

backward THz transmittance at different pump fluences with a TB at $t_d = 0$.

Optical pump-THz probe time-domain spectroscopy provides an excellent platform to directly investigate the dynamics of metasurface with the TB, which can be fully resolved by injecting free carriers on timescales much shorter than the THz signals. Figure 4a presents the microscopic image of the fabricated metasurface, with a single unit cell enlarged for better visualization. The mappings of the measured THz transmittance in the time domain, plotted as a function of $t_d$, are plotted in Fig. 4b,c for the forward and backward propagation, respectively. When measurements are conducted with the pump pulse arriving at the metasurface 3 ps before or after the THz probe pulse (that is, $|t_d| > 3\,\text{ps}$), a reciprocal outcome emerges near 0.7 THz, corresponding to the steady-state response of the metasurface. However, within the range $-3\,\text{ps} < t_d < 3\,\text{ps}$, a pronounced peak at 0.7 THz emerges, notably surpassing its counterpart in the backward spectrum in amplitude. The position where the TB occurs is denoted by white dashed lines ($t_d = 0$). Under the optical pump, the conductivity of the Ge film undergoes significant changes, leading to $t_d$-dependent variations in the nonradiative loss. The nonreciprocal polarization conversion process relies on the time derivative of the transition around the TB.

We further analyze the correlation between the transmittances and the pump fluences. With $t_d = 0$, we obtained the output transmittance as a function of the input power. Figure 4d presents the maximum transmission of THz through the metasurface. A nonlinear correlation between the transmittance and the input power indicates carriers' absorption saturation under the photoexcitation. The broadband THz transmittance spectrum of TB metasurface for forward and backward propagating is compared in Fig. 4e,f under the impact of varying optical pump fluences ranging from 0 to 2.5 $\text{mJ/cm}^2$. As the pump fluence increases, the intrinsic insulating region within the gap of SRR becomes progressively more conductive. This metamorphosis reconfigures the current-flow path within the unit cell, effectively extinguishing the fundamental mode of Ge-integrated SRR and precluding its coupling with other modes. At a pump fluence of 2.5 $\text{mJ/cm}^2$, we demonstrate pronounced nonreciprocal behavior with isolation exceeding 30 dB.

**Discussion**

We have experimentally demonstrated the TB-induced nonreciprocal transmission in the THz regime, which is completely different from the existing principles relying on a magnetic field, spatiotemporal modulation, or nonlinearity. The nonreciprocal polarization conversion is attributed to the transient loss variation, which is created by the controllable photogenerated carriers. It is also noteworthy that the isolation efficiency of the TB metasurface can be further improved and controlled by optimizing the metasurface design, engineering the width of the TB, or employing multilayer or reflective constructions[45]. The concept of TB can be readily extended to higher frequencies, such as infrared and visible light, as well as to other classical or quantum optical systems. Furthermore, introducing electrical tunability may push our nonreciprocal devices toward more practical miniaturized applications[46]. Our approach to realizing time boundaries using external stimuli can be configured to simultaneously introduce spatial and temporal boundaries through a spatial light modulator. Integrating spatial and temporal degrees of freedom enables even greater flexibility in real-time, reconfigurable wave control and manipulation. These findings pave a new avenue for developing innovative THz devices and provide opportunities in the rising field of time-varying photonic media, non-Hermitian physics, and Floquet physics in the presence of multiple input waves and multiple time interfaces, as in the case of time crystals[47, 48].

**Acknowledgments:**

Key Research and Development Program of the Ministry of Science and Technology grant 2022YFA1405200 (Y.Y.)

Key Research and Development Program of the Ministry of Science and Technology grant 2022YFA1404900 (Y.Y.)

Key Research and Development Program of the Ministry of Science and Technology grant 2022YFA1404704 (H.C.)

Key Research and Development Program of the Ministry of Science and Technology grant 2022YFA1404902 (H.C.)

National Natural Science Foundation of China (NNSFC) grant 62175215 (Y.Y.)

National Natural Science Foundation of China (NNSFC) grant 61975176 (H.C.)

National Natural Science Foundation of China (NNSFC) grant 62305384 (Y.H.)

National Natural Science Foundation of China (NNSFC) grant 62305298 (M.T.)

China National Postdoctoral Program for Innovative Talents grant BX20230310 (M.T.)

Youth Innovation Talent Incubation Foundation of National University of Defense Technology grant 2023-lxy-fhij-007 (Y.H.)

Key Research and Development Program of Zhejiang Province grant 2022C01036 (H.C.)

Fundamental Research Funds for the Central Universities grant 2021FZZX001-19 (Y.Y.)

Excellent Young Scientists Fund Program (Overseas) of China (Y.Y.)

**Author contributions**

Y.Y., T.J., Y.H., and M.T. created the design. Y.H., M.T., and Y.Y. designed the experiment. Y.H. and M.T. fabricated samples. S.H. measured with assistance from Y.H. M.T. analyzed the data. Y.H. performed simulations, T.J., Y.Y., and H.C.

provided the theoretical explanations. M.T., Y.H. wrote the manuscript with input from Y.Y., T.J., and H.C. Y.Y. supervised the project. All authors contributed extensively to this work.

**Competing interests**

The authors declare no competing interests.

**Data and materials availability**

The data that support the findings of this study are available from the corresponding author upon reasonable request.

Nature Communications **10**, 1345 (2019).